\documentclass[12pt,preprint]{aastex}
\usepackage{graphicx}
\usepackage{epsfig}

\def\a4{\hsize 17.0cm \vsize 25.cm}
\newcommand{\der}[2]  { \frac{{\rm d}#1}{{\rm d}#2} }

\newcommand{\dif}     {{\rm d}}

\shorttitle{On the hydrodynamics of the matter reinserted within SSCs}
\shortauthors{Tenorio-Tagle et al.}

\begin{document}

\title{On the hydrodynamics of the matter reinserted within superstellar 
clusters}

\author{
Guillermo Tenorio-Tagle
\affil{Instituto Nacional de Astrof\'\i sica Optica y
Electr\'onica, AP 51, 72000 Puebla, M\'exico; gtt@inaoep.mx}
Richard W\"unsch 
\affil{Astronomical Institute, Academy of Sciences of the Czech 
Republic, Bo\v{c}n\'\i\ II 1401, 141 31 Prague, Czech Republic;
richard.wunsch@matfyz.cz} 
Sergiy Silich
\affil{Instituto Nacional de Astrof\'\i sica Optica y
Electr\'onica, AP 51, 72000 Puebla, M\'exico; silich@inaoep.mx}
\and 
Jan Palou\v{s}
\affil{Astronomical Institute, Academy of Sciences of the Czech
Republic, Bo\v{c}n\'\i\ II 1401, 141 31 Prague, Czech Republic;
palous@ig.cas.cz}}

\begin{abstract}We present semi-analytical and numerical models, accounting 
for the impact of radiative cooling on the hydrodynamics of the matter 
reinserted as strong stellar winds and supernovae 
within the volume occupied by young, massive and compact superstellar clusters.
First of all we corroborate  the location of the
threshold line  in the mechanical energy input 
rate vs the cluster size plane, found by Silich et al. (2004). Such a line 
separates clusters able to drive a quasi-adiabatic or a strongly radiative 
wind from clusters in which catastrophic cooling occurs within the star 
cluster volume. Then we show that  the latter, clusters above the threshold 
line, undergo a bimodal behavior in which the central densest zones cool 
rapidly and accumulate the injected matter to eventually feed further 
generations of star formation, while the outer zones are still able to drive 
a stationary wind. The results are presented into a series of universal 
dimensionless diagrams from which one can infer: the size of the two zones, 
the fraction of the deposited mass that goes into each of them and the 
luminosity of the resultant winds, for clusters of all sizes and energy 
input rates, regardless the assumed adiabatic terminal speed ($v_{A\infty}$). 
\end{abstract}


\section{Introduction}

Upon the formation of a massive and compact superstellar cluster (SSC) the 
continuous amount of energy deposited by the massive sources, through winds 
and supernova explosions (SN), causes a large overpressure. This results from 
an efficient thermalization of the deposited kinetic energy through their 
random collisions. In the adiabatic solution of Chevalier $\&$ Clegg (1985; 
see also the recent numerical calculations of Canto et al. 2000; Raga et al. 2001  and of Rockefeller et al. 2004, 2005)
the overpressure leads to an almost uniform temperature and 
density distribution within  the star cluster volume causing however a slight 
pressure gradient that allows  the gas velocity to grow from zero km s$^{-1}$ 
at the center to the sound speed ($a_s$) at the cluster edge ($r = R_{SC}$). 
The hot gas then freely streams to conform a stationary wind in which the 
mass input rate $\dot M_{SC}$ equals the amount of matter leaving the cluster 
($\dot M_{SC} = 4 \pi R_{SC}^2 \rho_{SC} a_s$), and the density ($\rho_w$), 
temperature ($T_w$) and pressure ($P_w$) rapidly approach their asymptotic 
trends ($\rho_w \propto r^{-2}$, $T_w \propto r^{-4/3}$ and 
$P_w \propto r^{-10/3}$) while the velocity approaches its terminal speed 
$v_{A\infty} = (2L_{SC}/\dot{M}_{SC})^{0.5} \approx 2a_s$; where $L_{SC}$ 
is the energy deposition rate. 

More recently it has been shown that the winds produced by massive 
and compact clusters rather than behaving adiabatically they become strongly 
radiative (see Silich et al. 2003, 2004). Radiative
cooling hardly affects the density and velocity distributions however it 
causes the temperature to plummet to $T\sim 10^4$ K at short distances 
from the star cluster boundary. This diminishes drastically the sizes of the 
associated X-ray envelopes predicted by the adiabatic solution. 
We have also shown that in the $L_{SC}$ vs $ R_{SC}$ plane there is a 
threshold line above which radiative cooling becomes catastrophic within the 
star cluster volume and this inhibits the development of a stationary wind 
(see Silich et al. 2004). For these cases two solutions were proposed (see 
Tenorio-Tagle et al. 2005a, b). Both of them depart from the fact 
that as matter cannot leave the cluster it would have to accumulate within 
the SSC volume. One of the solutions assumes that photoionization keeps the 
temperature $\sim 10^4$ K and in the absence of gravity accumulation would 
follow until $\dot M_{SC} = 4 \pi R_{SC}^2 \rho_{SC} c_{HII}$ and a  
dense isothermal wind can develop. The speed achieved in these winds is 
several times the sound velocity ($c_{HII}$), however smaller than the 
escape speed for compact and massive clusters  ($v_{esc} = 
(2 G M_{SC}/ R_{SC})^{0.5}$). Thus the true physical stationary solution for 
clusters above the threshold line appears when matter accumulation is 
balanced by gravitational instabilities which lead to further stellar 
generations and thus to positive feedback within the star cluster volume, 
when  matter  reinserted by massive stars is driven into star formation.  

The location of the threshold line has also been shown to be a strong function
of the metalicity of the matter reinserted by winds and supernovae 
(Tenorio-Tagle et al 2005b) as this strongly enhances the cooling rate. It 
depends also on the heating efficiency (see Silich et al. 2006),
or the amount of energy that after full thermalization of the ejecta is not 
immediately radiated away, as in the case of close neighboring sources, and  
thus it can be evenly spread within the star cluster volume, causing the
central overpressure (Stevens $\&$ Hartwell 2003, Melioli \& de Gouveia Dal 
Pino  2004).    
 
Here in section 2 we confirm, through an independent code, the location of 
the threshold line and then show that this is in fact not the whole story. 
Clusters above the threshold line are here shown to undergo a bimodal 
behavior in which radiative cooling leads to mass 
accumulation and further star formation in their central densest regions 
while the  outer zones are still able to drive a stationary wind. 
Section 2 also evaluates the 
strength of the resultant winds for clusters above the threshold line as well 
as the expected rate of star formation within their cool 
interiors. Section 3 
shows some of our numerical simulations. These fully confirm the 
results obtained through our semi-analytical code. Section 4 summarizes our results and gives our conclusions.

\section{The semi-analytic approach}

The hydrodynamic equations for the flow between the stagnation point 
($R_{st}$) and the edge of the cluster (see Silich et al. 2004) are:
\begin{eqnarray}
      \label{eq.1a}
      & & \hspace{-1.0cm}
\frac{1}{r^2} \der{}{r}\left(\rho_w u_w r^2\right) = q_m ,
      \\[0.2cm]
      \label{eq.1b}
      & & \hspace{-1.0cm}
\rho_w u_w \der{u_w}{r} = - \der{P_w}{r} - q_m u_w,
      \\[0.2cm]
     \label{eq.1c}
      & & \hspace{-1.0cm}
\frac{1}{r^2} \der{}{r}{\left[\rho_w u_w r^2 \left(\frac{u_w^2}{2} +
\frac{\gamma}{\gamma - 1} \frac{P_w}{\rho_w}\right)\right]} = q_e - Q,
\end{eqnarray}
where $u_w, \rho_w, P_w$ are the velocity, the density and the thermal
pressure of the thermalized matter, $q_m$ and $q_e$ are the mass
and the energy deposition rates per unit volume, respectively, and
$Q = n_e n_i \Lambda(T,Z)$ is the cooling rate. We use the equilibrium
cooling function, $\Lambda(T,Z)$, for optically thin plasma whose
temperature is $T$ and metallicity is $Z$ obtained by Plewa (1995). In all 
calculations the metallicity of the plasma was assumed to be solar.

An important difference arises when one considers massive star clusters, 
those with a large mechanical energy input rate, which  drain immediately a 
large fraction of the deposited energy
through  catastrophic cooling. This happens first within their central, 
densest regions and results in an immediate loss of pressure and of the outward
pressure gradient. Solutions for  such cases require that   the stagnation 
point ($R_{st}$; the point where the expansion
velocity, $u_w = 0$ km s$^{-1}$) is shifted from the central position $r = 0$
to a distance, $0  < R_{st} < R_{SC}$, from the star cluster center.
In such cases, the solution of equations (\ref{eq.1a} - \ref{eq.1c})
differs from that for low mass clusters. Indeed, in
the catastrophic cooling regime with $R_{st} \ne 0$, the constant in
the integral form of the mass conservation equation,
\begin{equation}
      \label{eq.1d}
\rho_w u_w r^2 = \frac{q_m r^3}{3} + C ,
\end{equation}
is not equal to zero. In this, more general case, it depends on the 
location of the stagnation point, $C = - q_m R^3_{st} / 3$, as at
the stagnation point the expansion velocity is zero, $u_w = 0$. The 
mass conservation equation (\ref{eq.1a}) is then reduced to
\begin{equation}
      \label{eq.2a}
\rho_w = \frac{q_m r}{3 u_w} \left(1 - \frac{R^3_{st}}{r^3}\right) .
\end{equation}
Using this equation one can replace terms $\rho_w u_w$ and $\rho_w u_w r^2$
in equations (\ref{eq.1b}) and (\ref{eq.1c}). Taking the derivative in 
equation (\ref{eq.1c}) and then replacing $\dif{P_w}/\dif{r}$ from equation 
(\ref{eq.1b}), we obtain
\begin{eqnarray}
\label{eq.2b}
      & & \hspace{-1.1cm}
\der{u_w}{r}  = \frac{1}{\rho_w} \frac{(\gamma-1)(q_e - Q) + 
              q_m \left[\frac{\gamma+1}{2}u_w^2 - \frac{2}{3}
              \left(1 - \frac{R^3_{st}}{r^3}\right) a_s^2\right]}
              {a_s^2 - u_w^2} ,
      \\[0.2cm]     \label{eq.2c}
      & & \hspace{-1.1cm}
\der{P_w}{r} = - q_m \left[\frac{r}{3}\left(1 - \frac{R^3_{st}}{r^3}\right)
                 \der{u_w}{r} + u_w\right] ,
\end{eqnarray}
where $a_s = (\gamma P_w / \rho_w)^{1/2}$ is the speed of sound.
Note, that equations (\ref{eq.2a} - \ref{eq.2c}) become identical with
equations (7 - 9) from Silich et al. (2004), if $R_{st} = 0$.
Note also that the relation between the gas number density 
and the temperature at the stagnation point remains the same for all cases 
(see Silich et al. 2004):
\begin{equation}
      \label{eq.3}
n_{st} = \left[\frac{q_e - q_m a^2_{st} / (\gamma - 1)}
       {\Lambda(T_{st},Z)}\right]^{1/2} , 
\end{equation}
where $a_{st}$ is the sound speed at $R_{st}$. Equation (\ref{eq.3}) 
indicates that strong radiative cooling reduces the temperature at the 
stagnation point below the adiabatic value, $T_A$, which can be derived
from the same equation if one assumes that the cooling rate,
$n^2_{st} \Lambda \to 0$. 

\begin{figure}[htbp]
\plotone{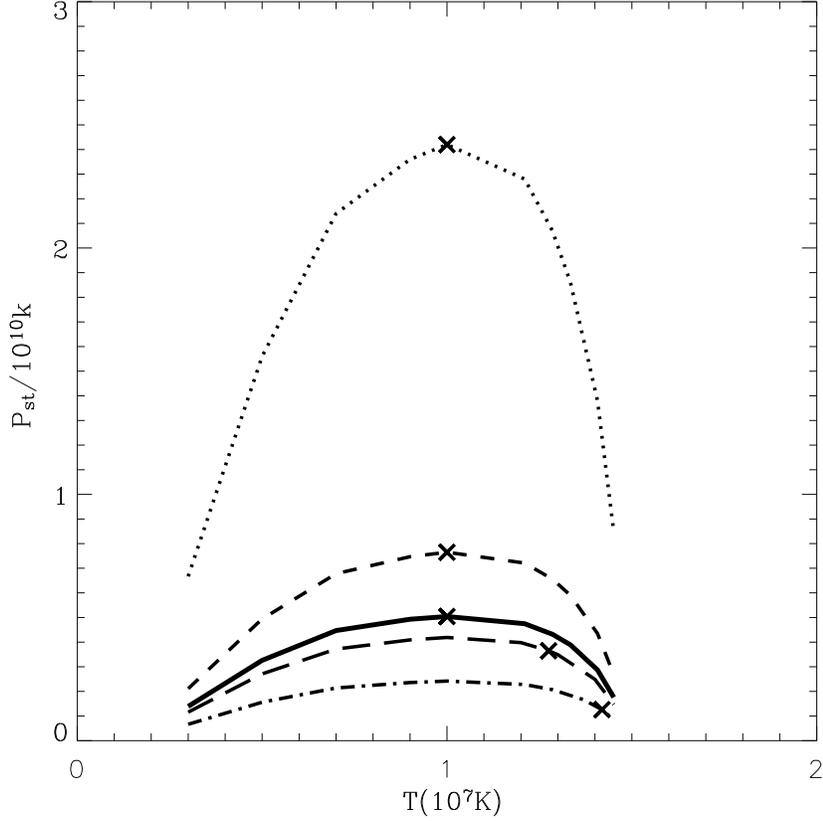}
\caption{Pressure and temperature at the stagnation point.  
The pressure at the stagnation point as a function of the gas
temperature for clusters with an $R_{SC}$ = 10 pc and different mechanical 
luminosities. The stationary wind solution requires a unique pressure for 
every cluster indicated here by X symbols. Note that $P_{st}$ remains 
below the maximum possible value for clusters whose $L_{SC}$ is smaller
than the threshold  mechanical luminosity (see section 2.2) and 
acquire the maximum value, $P_{max}$, when $L_{SC} \geq L_{crit}$.  The 
temperature at the stagnation point, $T_{st}$, for clusters above the 
threshold line, does not change. It is, $T_{st} \approx 10^7$K, if
$V_{A\infty} = 1000$ km s$^{-1}$. Each curve corresponds to a selected
$L_{SC}$ value. The case corresponding to the critical energy input rate,
$L_{crit} = 4.35 \times 10^{41}$ erg s$^{-1}$, is marked by the solid
line. Dashed and dotted lines are for clusters above the threshold line
with $L_{SC} = 10^{42}$ and  10$^{43}$ erg s$^{-1}$, respectively. The 
dash-dotted line and the long dashed line are for  clusters below the threshold
line with $L_{SC} = 10^{41}$ and 3 $\times 10^{41}$ erg s$^{-1}$, 
respectively.}
\label{fig1}
\end{figure}

\subsection{The threshold line}

For low mass clusters,  the stagnation point (which presents the largest 
pressure within the resultant distribution) ought to be  at the center and 
values of temperature and density of the plasma at such location may be 
derived  through equation (\ref{eq.3}) by iterating over $T_{st}$ until 
the sonic point ($R_{sonic}$) is accommodated at the star cluster 
surface, $R_{sonic} = R_{SC}$, as described in Silich et al. (2004). 
One can then calculate the pressure at the stagnation point (the central
pressure in these cases):
\begin{equation}
      \label{eq.4}
P_{st} = k T_{st} \left[\frac{q_e - q_m a^2_{st} / (\gamma - 1)}
       {\Lambda(T_{st},Z)}\right]^{1/2} , 
\end{equation}
and then integrate numerically equations (\ref{eq.2a}) - (\ref{eq.2c}) 
outwards from $R_{st} = 0$.

As shown in Figure 1 for a 10 pc cluster with an energy input rate equal to 
10$^{41}$ erg s$^{-1}$ and an assumed $v_{A\infty}$ = 1000 km s$^{-1}$ (lowest 
curve), the pressure value at $R_{st}$ that sets the sonic point at the 
cluster surface, is smaller than the maximum pressure $P_{max}$ that results 
from all other different values of $T_{st}$. Note however, that $P_{st}$ 
becomes larger  for larger clusters, approaching continuously the maximum 
allowed pressure (see equation \ref{eq.4} and Figure 1). Such a behavior 
occurs until the power of the star cluster, $L_{SC}$, reaches the threshold 
value $L_{crit}$ (see Figure 2). For such a cluster the pressure at the 
stagnation point acquires the 
maximum possible value from those allowed by the parameters of the cluster, 
$P_{st} = P_{max}$ (see Figure 1, solid line). This  selects the 
temperature (see Figure 1) and the density (equation \ref{eq.3}) at the
stagnation point and equations \ref{eq.2a} - \ref{eq.2c} may then be solved
numerically. For even larger energies, the 
selected $P_{st}$ has always its largest possible value and $T_{st}$ remains 
equal to the value acquired at the threshold line. However above the 
threshold line this is not sufficient to warrant  the location of the sonic 
point at the star cluster surface and the only possibility to establish a 
stationary outflow arises from displacing the stagnation point from 
the star cluster center.

\subsection{Solutions above the threshold line}

Figure 2 displays the location of the threshold limit as a function of 
energy and the cluster size for two different values of $V_{A\infty}$ (1000 
and 1500 km s$^{-1}$). Thus for clusters at and above the threshold line the 
pressure as a function of the temperature at the stagnation point is to 
have the maximum value, $P_{max}$, which occurs always at the same 
temperature for a given ratio of
$L_{SC} / {\dot M}_{SC} = V^2_{A\infty} / 2$  (see Figure 1).
One can then find the parameters of the injected plasma at the stagnation 
point by searching through a range of temperatures ($T_{st}$), smaller than 
the central temperature derived from the adiabatic solution, until the
pressure at the stagnation point has its maximum possible value. Having 
$P_{st}$, $T_{st}$, and $\rho_{st}$ one can 
solve equations \ref{eq.2a} - \ref{eq.2c} numerically for different
positions of the stagnation point. The proper value of the stagnation point 
radius is then selected by the condition that the sonic point lies at the
star cluster surface. As shown in Figure 1, for clusters above the threshold 
line, this procedure selects among a broad spectrum of formal solutions, all 
of them valid integral  curves, the unique single valued solution. 
This places, for a given set of star cluster parameters, $R_{st}$ at the 
closest possible distance to the center of the cluster. 

Figure 3 shows how the fractional radius ($R_{st}/R_{SC}$) changes when the 
selected normalized cluster luminosity, $L_{SC}/L_{crit}$, is larger than 1. 
Note that this relationship is  universal for clusters of all sizes and 
masses if one selects the value of $L_{crit}$ that corresponds to the assumed 
value of $V_{A\infty}$. The location of $R_{st}$ asymptotically approaches 
$R_{SC}$ for increasing $L_\mathrm{SC}$ values, implying that the outer 
cluster regions are to eject the deposited matter in the form of a high 
velocity stationary wind even if the mechanical luminosity of the star 
cluster exceeds the critical value, $L_{SC} > L_{crit}$. This implies that 
the isothermal wind solution discussed in Tenorio-Tagle et al. (2005a), 
under the assumption of matter accumulation within the whole cluster
volume, is in reality inhibited by the outward expansion of the matter 
deposited between $R_{st}$ and the cluster surface.

\begin{figure}[htbp]
\plotone{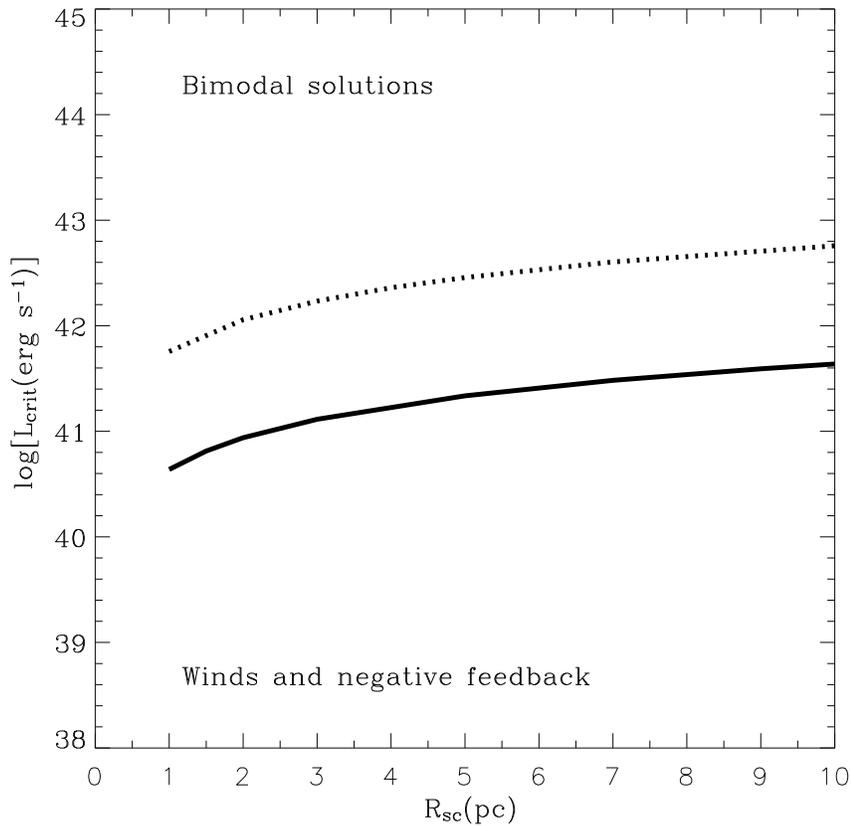}
\caption{The threshold line. The threshold line calculated under the assumption of a $V_{A\infty}$ equal to
1000 km s$^{-1}$ (solid line) and 1500 km s$^{-1}$ (dotted line).
The threshold  line separates two regions in the 
mechanical energy input rate or SSC mass, vs the cluster size plane. 
Clusters below the threshold line lead to stationary winds, either quasi 
adiabatic (far below the threshold line) or strongly radiative (as 
one approaches the threshold line). Clusters above the threshold line are 
here shown to produce a bimodal solution in which their densest inner regions 
radiate away  immediately  the deposited energy while the outer zones develop 
a strongly radiative stationary wind.} 
\label{fig2}
\end{figure}

\begin{figure}[htbp]
\plotone{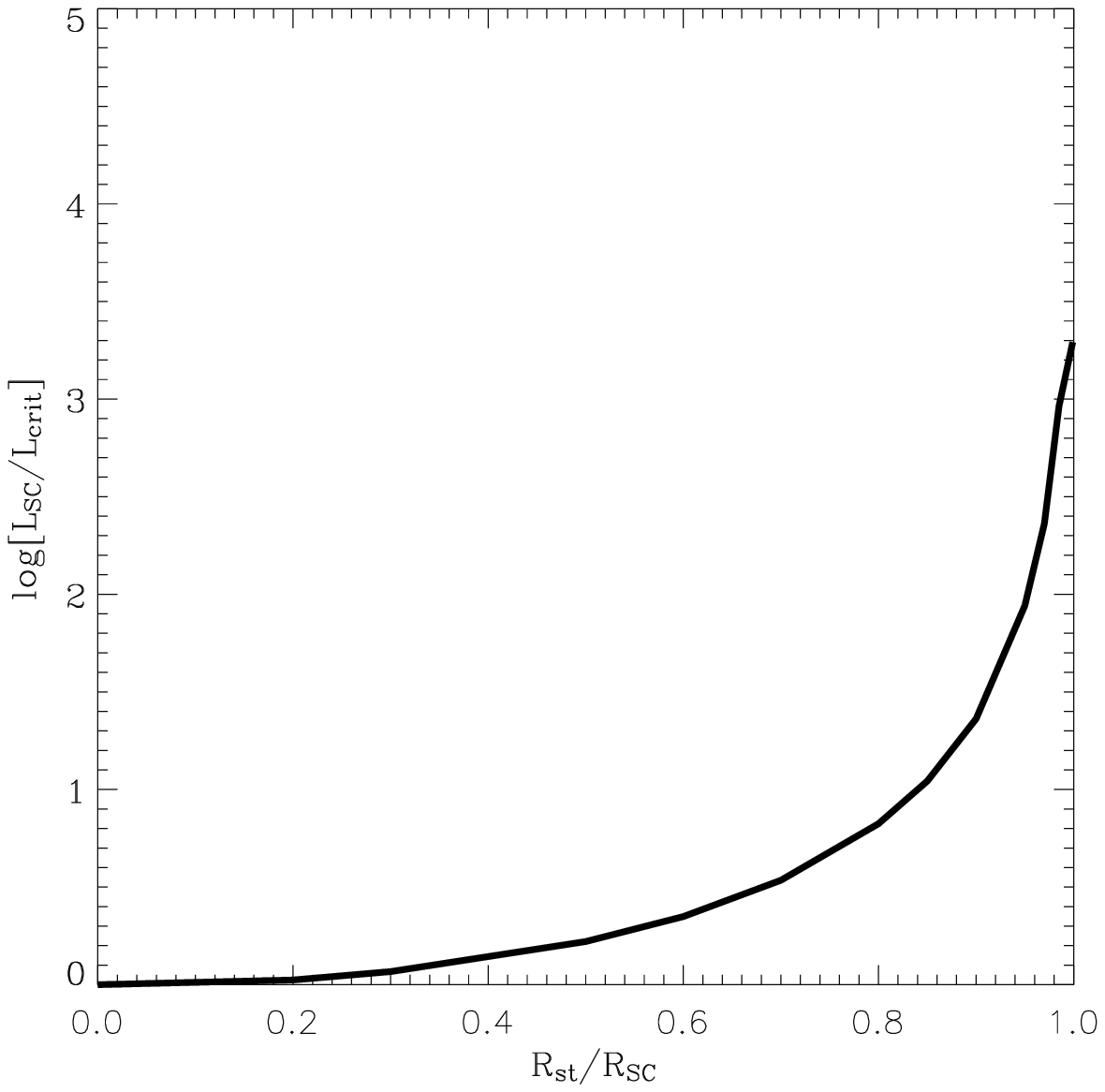}
\caption{The stagnation radius. For a given set of cluster parameters above 
the threshold line, the ratio of the mechanical energy input rate over the 
corresponding critical energy input rate (shown in Figure 2), define the 
location of the stagnation radius able to establish a stationary outflow.   
In all cases the  stagnation radius is located at the minimum possible 
distance from the cluster center.}
\label{fig3}
\end{figure}

\subsection{The bimodal hydrodynamic  behavior of clusters above the 
threshold line.}

Figure 4 shows the impact of cooling on the resultant winds. The resultant density 
profiles are hardly affected by cooling and thus rapidly approach an $r^{-2}$ 
distribution (see Figure 4). However, the further  
above the threshold line that a cluster may be, the larger  the impact
of cooling also on the resultant wind. The  temperature profile strongly 
deviates from the adiabatic model predictions ($T_w \propto r^{-4/3}$) 
already for clusters with an  $L_{SC}$ approaching the threshold line. For 
such clusters the temperature falls down and  rapidly reaches
$10^4$ K at some distance from the star cluster surface  
(Figure 4). The process is enhanced for clusters above the threshold line, 
causing the temperature to plummet down to $10^4$ K even closer to the 
cluster surface (see for example dotted and dashed lines in Figure 4). 

Above the threshold line  the fraction of the injected matter,
${\dot M}_{out} / {\dot M}_{SC}$, that clusters return to the
ISM of the host galaxy, decreases monotonically with the rate of
mechanical energy (see dotted line in the upper panel of Figure 5). The rest 
of the injected matter, ${\dot M}_{in} / {\dot M}_{SC}$, remains bound within
the stagnation radius (solid line in the upper panel of Figure 5) and thus is 
due to enhance its density, promoting an even faster cooling, as it 
accumulates. The accumulation process is so rapid that soon becomes also 
stationary, as $\dot M$ becomes equal to the star formation rate 
(see Tenorio-Tagle et al. 2005b). Both fractions 
(${\dot M}_{out} / {\dot M}_{SC}$ and ${\dot M}_{in} / {\dot M}_{SC}$) become 
universal functions for all clusters  if normalized to the corresponding 
threshold mechanical luminosity, $L_{crit}$, (see Figure 2).

At the same time, the flux of energy at the star cluster surface, $L_{out}$, 
normalized in a similar way, sharply decays for clusters with increasing
mechanical luminosity (Figure 5b). This implies that above the threshold line,
the cluster wind carries away only a fraction of the mechanical energy 
deposited by supernovae and stellar winds and this is smaller than that 
expected from starburst synthesis models (e.g. Leitherer et al. 1999).

Such powerful clusters, with a stagnation point approaching $R_{SC}$ 
(see section 2.2), have thus a cool interior and produce also a cool outflow, 
undetectable  in the X-ray regime. This  exposed to the UV 
radiation from the central cluster may be detected as a low 
intensity broad component in the emission spectra associated
with massive, compact star clusters. 
\begin{figure}[htbp]
\vspace{17.5cm}
\includegraphics{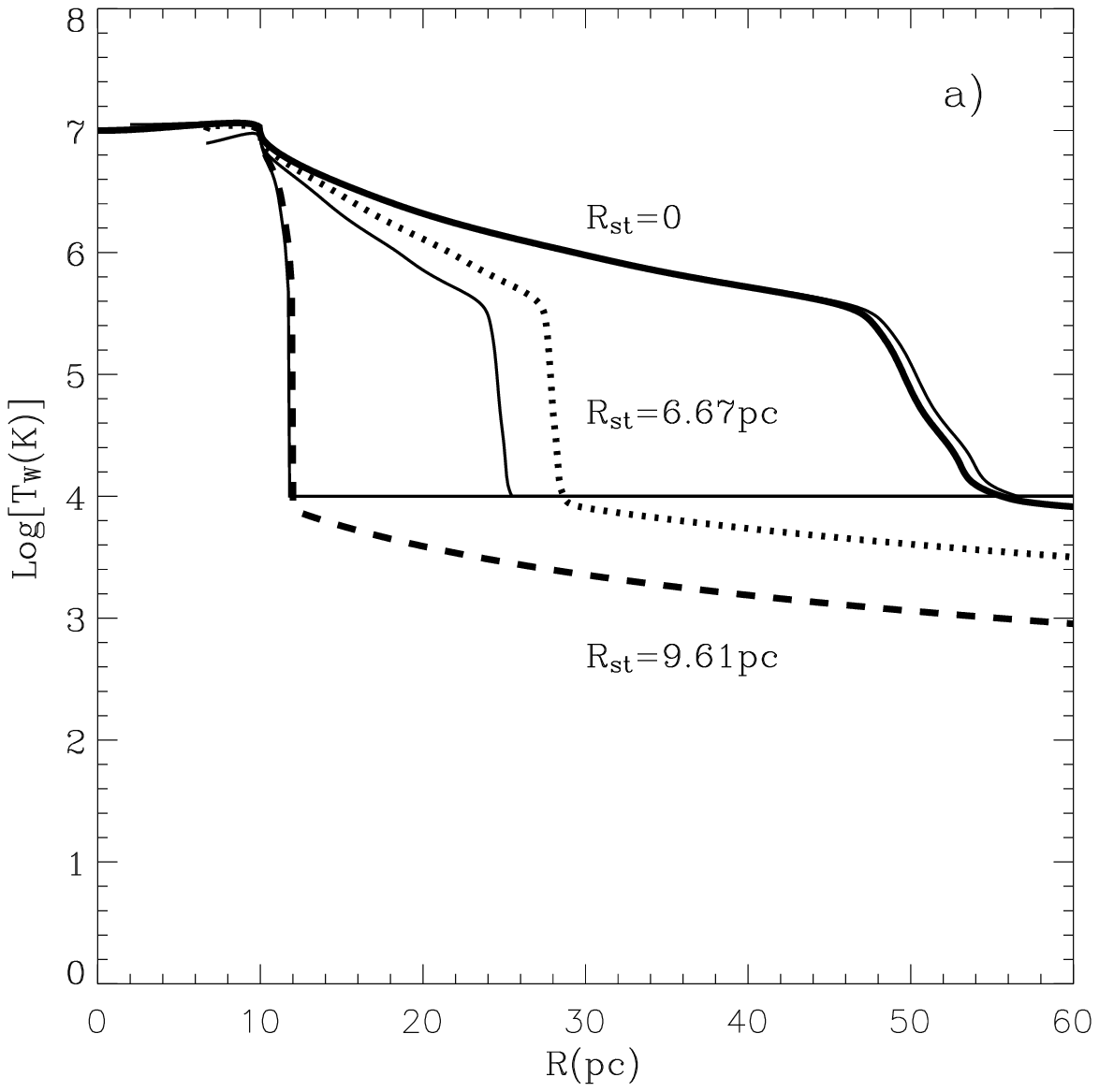}
\includegraphics{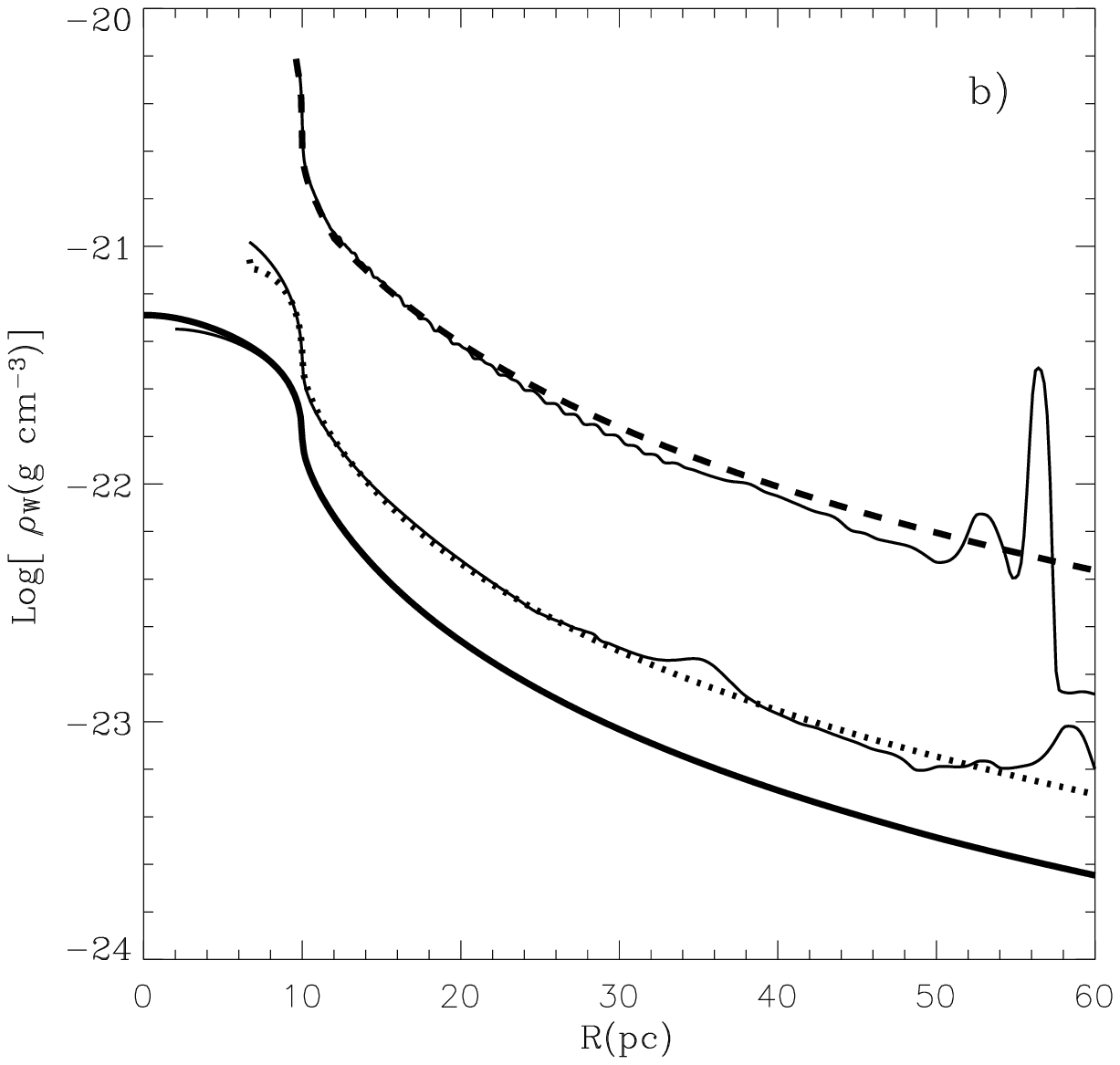}
\caption{The impact of strong radiative cooling on the resultant winds. 
Temperature (panel a) and density (panel b) profiles of the 
stationary winds developed by clusters with an $R_{SC} = $ 10 pc, and a  
mechanical energy input rate  equal to $4.35 \times 10^{41}$ erg s$^{-1}$,
$1.27 \times 10^{42}$  erg s$^{-1}$ and $6.2 \times 10^{43}$ erg s$^{-1}$ 
(solid, dotted and dashed  lines, respectively). Note that the less energetic 
cluster lies below the threshold line (see Figure 2) and it thus has its 
stagnation radius at the cluster center. The resultant stagnation radius 
for each case is indicated next to each of the solutions. Results from the 
semi-analytical method are shown by thicker lines, and are compared with 
the numerical results (see section 3) shown by the thin solid lines. 
In all numerical calculations, the lower temperature limit was set to 
$10^4$ K. 
Above the threshold line the flow is not absolutely stationary 
and as a result there are some oscillations of the stagnation point location, 
causing recurrent variations in the radius at which the temperature 
drops to 10$^4$ K. Another aspect of the non-stationary behavior are 
density enhancements visible in the outer parts of the wind.
}
\label{fig4}
\end{figure}

\begin{figure}[htbp]
\vspace{17.5cm}
\includegraphics{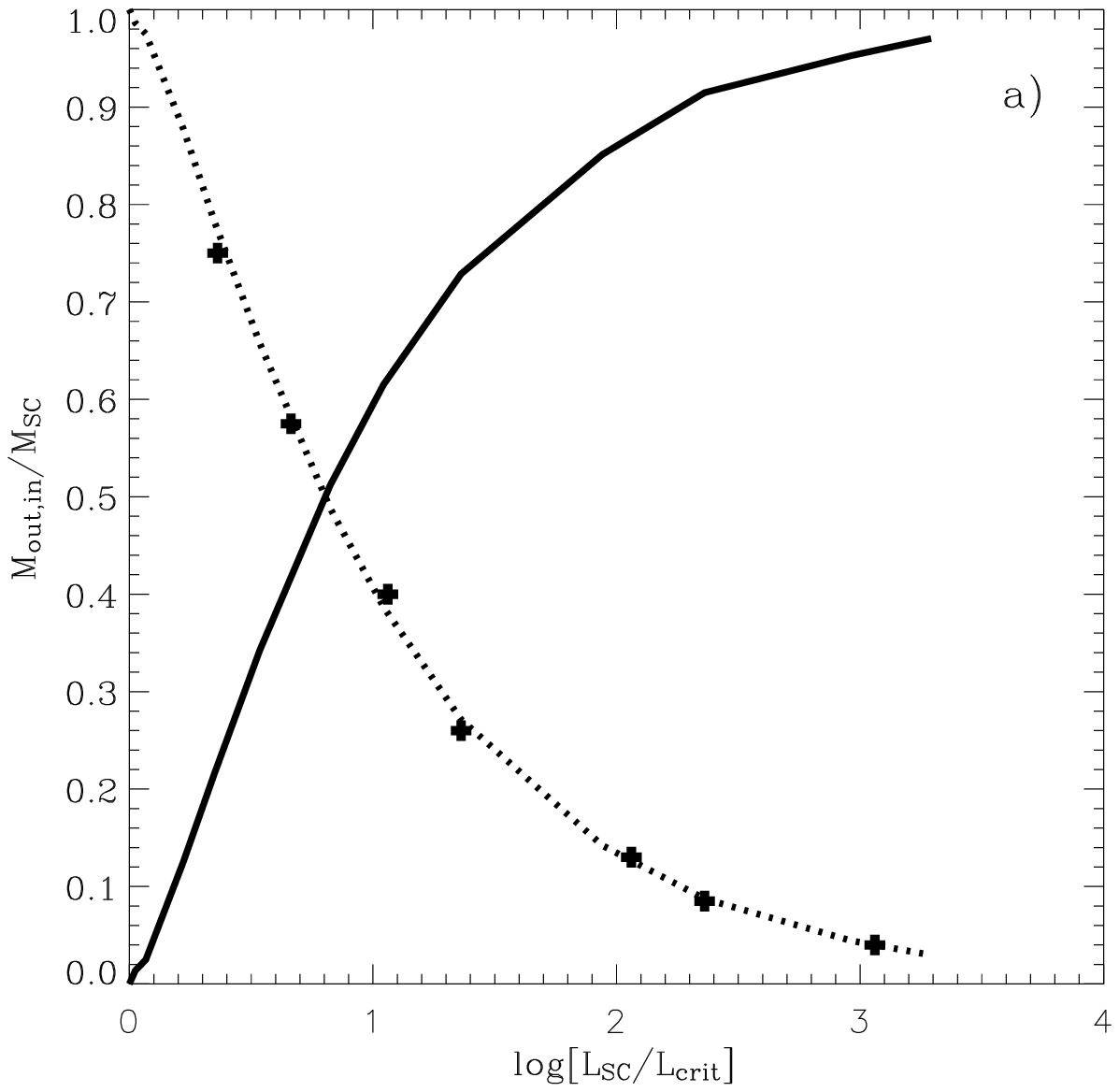}
\includegraphics{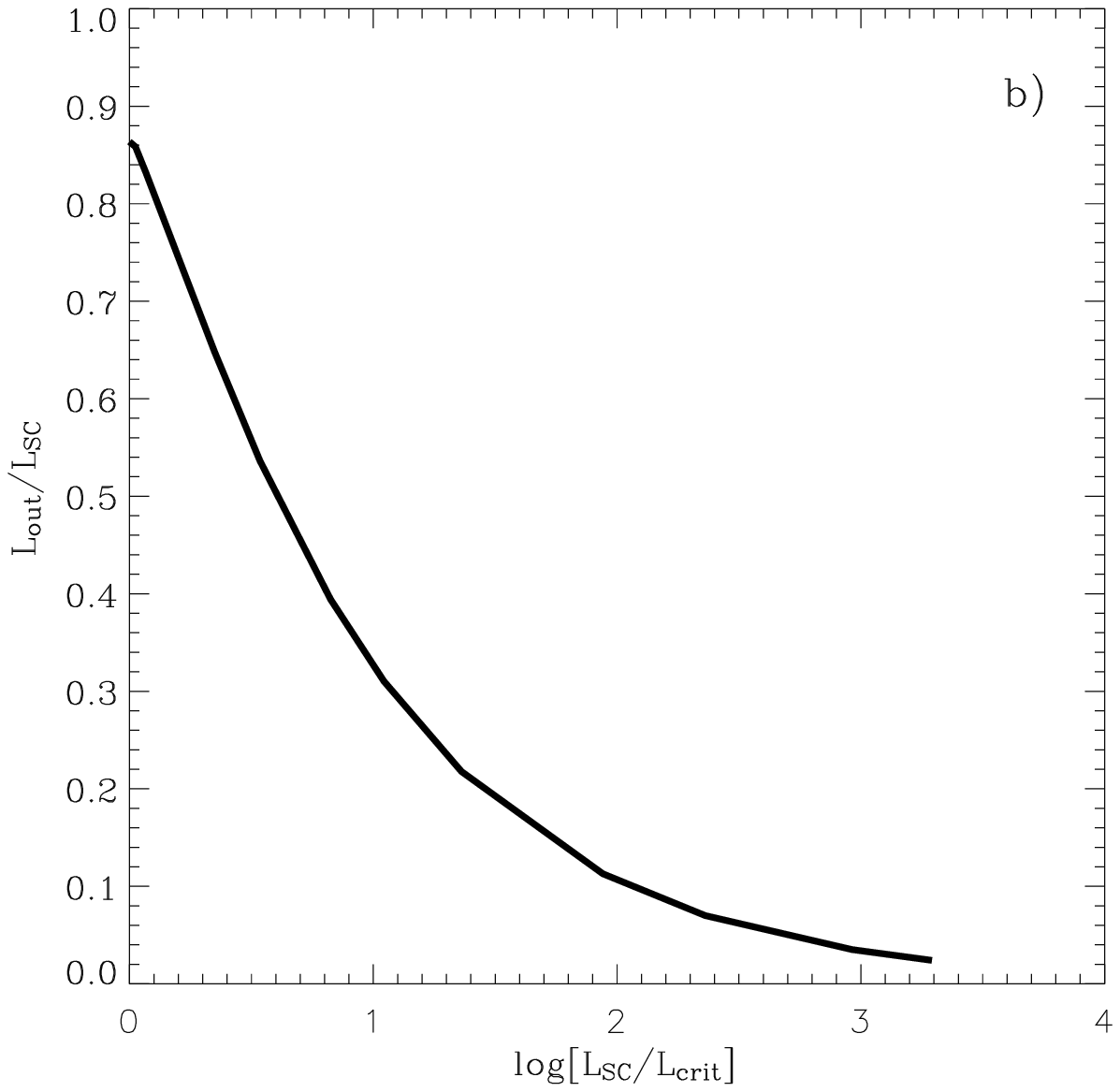}
\caption{Solutions above the threshold line. The normalized mass and energy 
output for star clusters whose mechanical luminosities exceeds the critical 
value. The dotted line in panel a) indicates the fraction of the 
deposited matter (${\dot M}_{out}$) that leaves the cluster as a stationary 
wind. This is compared with the results from the numerical simulations 
indicated by the cross symbols. The solid line in the panel a) displays the 
fraction of the deposited matter (${\dot M}_{in}$) that remains bound to 
the cluster and is thus accumulated within $R_{st}$ to become a source of  
secondary star formation. Panel b) shows that winds from clusters above 
the threshold line carry only a fraction of the energy deposited by 
supernovae and stellar winds within the cluster volume. Note that as 
Figure 3, both diagrams are applicable to clusters of all sizes and energy 
input rates once the appropriate $L_{crit}$, bound to the assumed 
$v_{A\infty}$, is selected.}
\label{fig5}
\end{figure}

\begin{figure}
\vspace{12.0cm}
\includegraphics{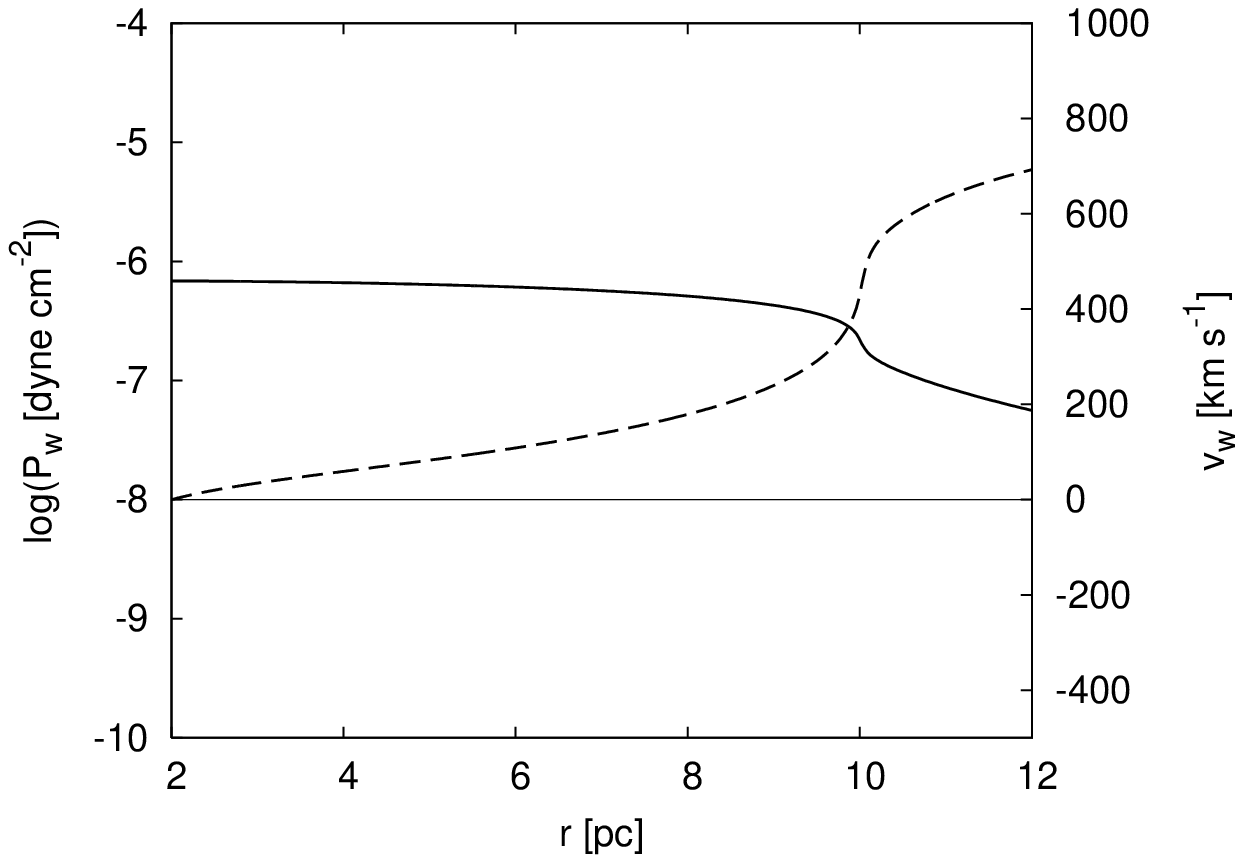}
\includegraphics{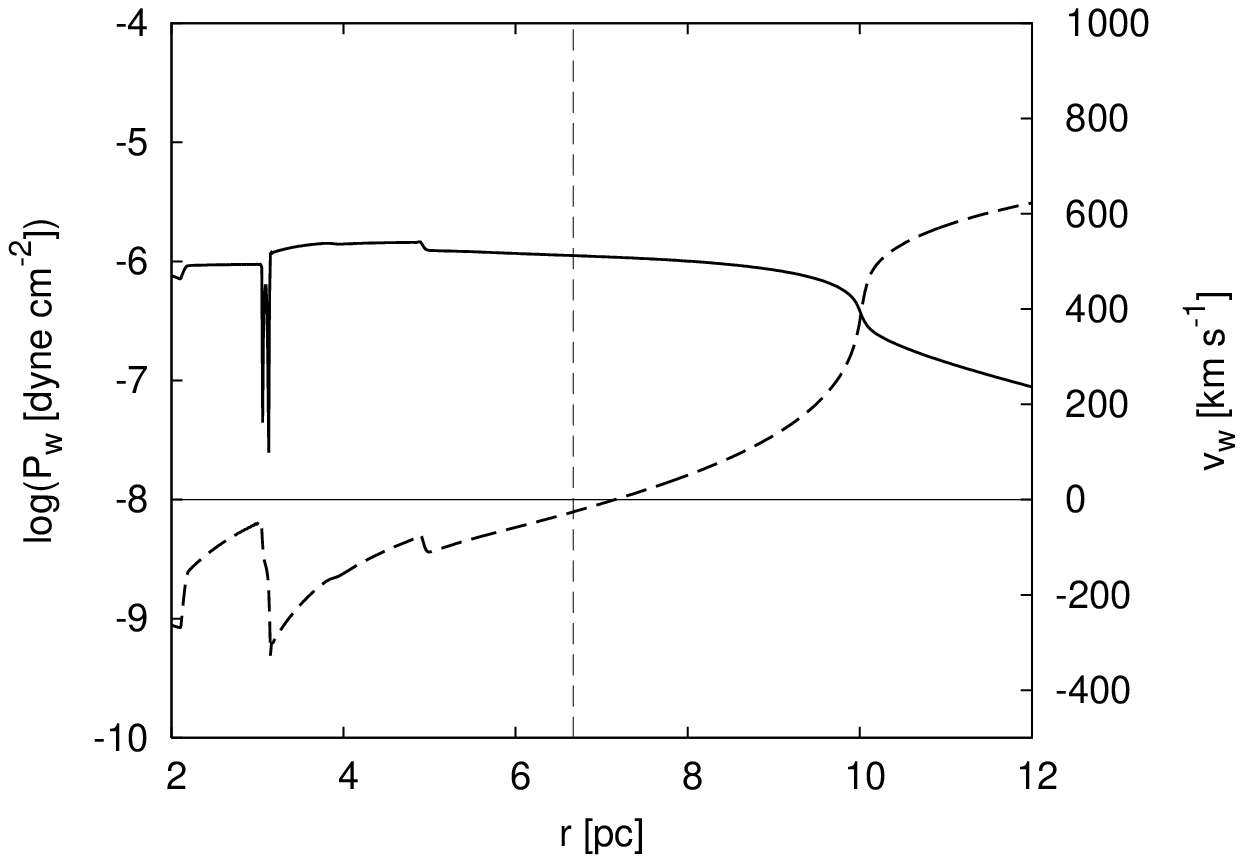}
\includegraphics{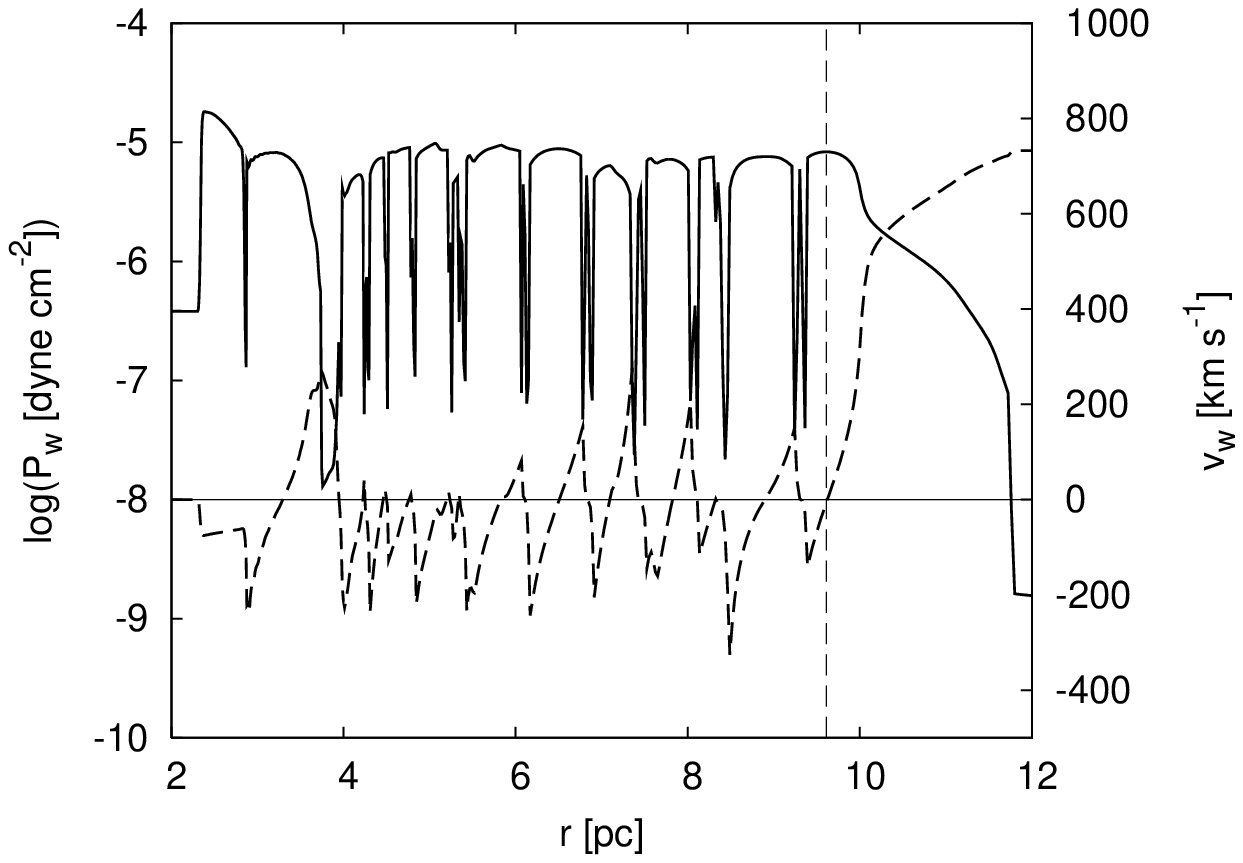}
\caption{The flow within the star cluster volume. Radial profiles of 
pressure (solid lines) and velocity (dashed lines) for the same cases 
presented in Figure 4. The horizontal lines mark the zero velocity and 
the vertical dotted lines the location of the stagnation point. Note 
that for the lowest energy case (upper left panel), the latter is at the 
center of the cluster. In the medium energy case (upper right panel), the 
repeated passage of the cooling front propagating rapidly from the cluster 
center outwards is followed by an inward motion of a shock front to the low
pressure region. This results in the oscillations of the stagnation
radius. In the high energy case (lower panel), the dense standing
shells form by collisions of shocks moving from both sides into the 
low pressure cold space located at some distance from the cluster center.
Note however that the position acquired by the stagnation point, allows 
the wind  in each case to attain its sonic velocity at the cluster surface.}
\label{cf}
\end{figure}

\section{The numerical approach}

All numerical calculations here presented are based on the finite difference 
hydrodynamic code ZEUS for which the cooling routine has been  reimplemented 
to make it suitable for the modeling of extremely fast cooling regions in 
the wind or within the SSC volume. Further, we solve the implicit form of the 
energy equation using the Brent algorithm, which is faster and more stable 
than the original Newton-Raphson method. Finally, we include the cooling 
rate on the computation of the time-step. The amount of energy which can be 
radiated away from a given cell during one time-step must be smaller
than 10\% of its internal energy. The time-step is decreased to meet this 
cooling rate condition. However, since this could lead to extremely 
small time-steps, what would substantially degrade the overall code 
performance, we do not allow the {\em global} time-step to decrease below 
0.1 times the "hydrodynamic" time-step determined by the
Courant-Friedrich-Levi criterion. In case a certian cell requires an even 
smaller time-step due to the cooling rate condition, we subdivide the
time-step even more. Using such small enough time-sub-steps, we numerically 
integrate the energy equation with the source terms only in the affected 
cell(s). This "time refinement" is applied only {\em locally}, so the CPU 
time is not wasted in cells where the high time resolution is not required.

The numerical models are in a good agreement with our semi-analytical code 
(see Silich et al. 2004). Figure 4  shows some examples from our 
many test calculations with different cluster parameters, compared with the 
semi-analytical results (bold face lines in Figure 4). The numerical models 
converge to the semi-analytical solution in runs with an  increasing
resolution. A higher resolution is more important in the
central region, and therefore we use a scaled grid (i.e. radial size of grid
cells $dr$ proportional to $r$). Simulations on a
scaled grid give results comparable to runs on an equidistant grid with
twice as many cells.

The resolution convergence was tested for 200, 400, 800, and
1200 radial grid cells for $L_{SC} = 3 \times 10^{41}$ erg s$^{-1}$. 
The results show a good agreement in the whole computational domain. 
The largest differences appear at the radius where the temperature
drops. For 200/400, 400/800, 800/1200 this region is radially shifted by 
2, 0.4 and 0.1 pc. In all other simulations we use the highest resolution, 
1200 grid cells.

\subsection{Boundary and initial conditions}

Within the star cluster volume, mass and energy per unit volume 
($q_m$ and $q_e$) are continously and homogeneously replenished 
($q_\mathrm{m} = (3\dot{M}_\mathrm{sc})/(4\pi
R_\mathrm{sc}^3)$ and $q_\mathrm{e} = (3L_\mathrm{sc})/(4\pi R_\mathrm{sc}^3)$)
using the following procedure at each time-step: 1) the
density and the total energy in a given cell are saved to $\rho_\mathrm{old}$ 
and $e_\mathrm{tot,old}$, 2) new mass is inserted $\rho_\mathrm{new} =
\rho_\mathrm{old} + q_\mathrm{m} dt$, the velocity is corrected so that the
momentum is conserved $\mathbf{v}_\mathrm{new} =
\mathbf{v}_\mathrm{old}\rho_\mathrm{old}/\rho_\mathrm{new}$, 3) internal energy
is corrected to conserve the total energy $e_\mathrm{i,mid} = e_\mathrm{tot,
old} - \rho_\mathrm{new}\mathbf{v}_\mathrm{new}/2$; 4) the new energy is 
inserted in a form of internal energy $e_\mathrm{i,new} = e_\mathrm{i,mid} +
q_\mathrm{e}dt$. The computational domain extents in a range $(R_\mathrm{in},
R_\mathrm{out})$, where $0<R_\mathrm{in}\ll R_\mathrm{sc}$. An open boundary
condition was applied at both ends of the computational grid, $R_\mathrm{in}$ 
and $R_\mathrm{out}$. The initial conditions for clusters below the threshold 
line were derived from  semi-analytical solutions. For clusters above the 
threshold line, a semi-analytical solution with the same $R_\mathrm{sc}$ and 
$v_\infty$ and a $L_\mathrm{sc}$ just below $L_\mathrm{crit}$ was used. This 
is then followed by a continuous replenishment of the appropriate  
$q_m$ and $q_e$ that correspond to the selected $L_{SC}$.

Here, as examples, we present three calculations for the mechanical energy 
input rate equals to $4.35 \times 10^{41}$ erg s$^{-1}$,
$1.27 \times 10^{42}$ erg s$^{-1}$ and $6.2 \times 10^{43}$ erg s$^{-1}$. 
The other parameters were $R_{SC} = 10$ pc, $V_{A\infty} = 1000$ km s$^{-1}$, 
$R_{in} = 2$ pc and $R_{out} = 100$ pc. The resolution was 1200 grid cells 
in the radial direction.

\subsection{1-D numerical simulations}

Our numerical scheme is able to reproduce with great accuracy the run of 
density, velocity and temperature obtained with the semi-analytical method  
for clusters below the threshold line (see Figure 4). For clusters above the 
threshold line the numerical simulations  become even more powerful as with 
such a method one is also able to calculate the flow everywhere 
within the cluster volume (see Figure 6) and not only in the region exterior 
to the stagnation point (as with the semi-analytical method, see section 2). 
Here, as examples, we present  three calculations for which $R_{SC}$ = 10 pc 
and the mechanical energy input rate  equals  
$4.35 \times 10^{41}$ erg s$^{-1}$, $1.27 \times 10^{42}$  erg s$^{-1}$ and 
$6.2 \times 10^{43}$ erg s$^{-1}$. Figure 4 displays the large-scale flow
and Figure 6 (pressure and velocity plots) provide details  within the star 
cluster volume.

The lowest energy case (close to the threshold line) clearly shows the 
stagnation point at the cluster  center, there where the pressure is 
largest, and the sonic point at the cluster surface (see Figure 6 upper left 
panel). For clusters above the threshold line (see upper right and lower 
panels in Figure 6), the central regions of the cluster volume cool 
rapidly down to $\sim 10^4$~K. 
This causes the highly non-stationary  behavior of the flow there, with a
number of the radiative shocks and cooling fronts passing through the inward 
moving material (see Figure 6). This results in multiple  local density 
enhancements. Some of these may acquire   positive velocities and move pass  
the stagnation point and perturb the quasi-stationary region between $R_{st}$ 
and $R_{SC}$. These perturbations are also visible in the outer parts of the 
wind (see Figure 4, panel b). Nevertheless, the free wind region, 
$r > R_{SC}$, remains quasi-stationary and its inner structure is found to be 
in good agreement with that predicted in the semi-analytical calculations 
(see Figure 4). 
 
Thus the numerical calculations, in agreement with the semi-analytical 
results, basically confirm the bimodal behavior for clusters above the 
threshold line. For such clusters, the outer region  with $r > R_{st}$ 
presents a positive increasing velocity, while in the 
cold interior, with $r < R_\mathrm{st}$, the velocity $u_w$ may 
present a complicated time dependent profile associated to the passage of 
rapid cooling fronts over the stagnant matter. 
The mass inserted into the outer region leaves the cluster as a
stationary wind, while the mass inserted into the cold inner zone
partly accumulates there, partly flows into the center through the
inner boundary, or partly enters the region outside the $R_{st}$ and 
perturbs the wind at $r > R_{st}$ (see Figures 4 and 6).
For the largest mechanical energy input rate, multiple high 
density and low temperature shells form within the volume defined by the 
stagnation point, $r < R_{st}$. These result from a rapid  mass 
inflow from both their sides, which makes them  grow in mass as likely 
seeds  for further star formation.

\section{Conclusions}

We have presented here semi-analytical and numerical simulation of the 
hydrodynamics undertaken by the matter reinserted via winds and supernovae 
by the  sources evolving within the volume occupied by a superstellar cluster.
All of these calculations are based on the assumptions made originally
in the adiabatic solution of Chevalier \& Clegg (1985) regarding an even 
spacing between sources, a full conversion of the kinetic energy injected 
within the star cluster volume into thermal energy and a smooth distribution 
of the injected material. These assumptions lead to a gas 
with a sound speed at the cluster edge much larger than the escape speed, 
even in the case of very massive and compact clusters, and thus gravitational 
effects have been ignored by all workers in the field.

Our results, that cover the full parameter space in the mechanical energy 
deposition rate (or mass of the SSC) vs the cluster size ($L_{SC}$ vs 
$R_{SC}$) plane, account for radiative cooling.  We have confirmed 
with 1D numerical simulations the 
location of the threshold line on the $L_{SC}$ vs $R_{SC}$ plane 
(see Silich et al. 2004). Solutions far below the threshold line  lead to 
quasi-adiabatic stationary winds, while more massive clusters, approaching 
the threshold line, produce stationary strongly radiative winds for which 
their temperature departs from the adiabatic solution and drops to 10$^4$ K 
close to the star cluster surface. In all of these cases  the stagnation 
point remains at the cluster center and the sonic point is at the cluster edge.
The iteration over temperature then defines the appropriate initial 
conditions and allows to integrate the basic equations numerically.

The main focus of this paper however has been on the solution for clusters 
above the threshold line and our results superseed some of our previous 
findings. In particular, the possibility of an isothermal wind, or 
supernebula, calculated in Tenorio-Tagle et al. (2005a) under the assumption 
of mass accumulation everywhere within the star cluster volume, has here been 
shown to be not possible. SSCs above the threshold line lead instead to  an 
intrinsic bimodal behavior: Their densest inner regions do cool immediately, 
depleting the pressure and outward pressure gradient required to drive an 
outflow, while their outer zones
manage, in all cases, to compensate radiative cooling with the energy input 
rate and thus remain hot and able to establish a stationary wind. In all of 
these cases the fraction of the cluster volume that cools down and enters a 
phase of matter accumulation and further star formation, as well as the 
fraction that drives a stationary wind, is decided by the location adopted 
by the stagnation point. 
 
The value of $R_{st}$ is the minimum radius above which the stationary
solution exists. Within $R_{st}$ cooling forms low pressure regions into 
which the inserted mass flows. The flow can be either oscillatory (medium
energy case) or colliding (high energy case) forming dense shells out 
of mass arriving from both sides (see figure 6).
In the semi-analytic approach the location of the stagnation radius within 
the cluster volume is uniquely defined by the appropriate $L_{crit}$ 
corresponding to the assumed $v_{A\infty}$, and the star cluster mechanical 
luminosity, $L_{SC}$. This, at the same time, uniquely defines the amount 
of matter and the energy flux injected as a stationary wind as well as the 
amount that accumulates within the cluster inner regions and which eventually 
is driven into further star formation. Our diagrams (Figures 3 and 5) are 
thus applicable to clusters of all sizes and all possible energy input rates 
as long as the mechanical energy is scaled with the appropriate $L_{crit}$, 
corresponding to the assumed $v_{A\infty}$.

A further implication of clusters above the threshold line is that their 
mechanical energy and mass input rates into their surroundings are 
considerably smaller than the values predicted from synthesis models of 
coeval clusters. Cooling and recombination within the stagnation radius of 
such clusters would also reduce considerably their UV photon output.

Further calculations in two dimensions accounting for radiative cooling
and also for the self-gravity of the cluster, are now underway and would
be the subject of a future communication. In these calculations, thermal
instabilities lead to the formation of dense clumps often surrounded by a 
less dense hot wind. Preliminary results indicate that some of these clumps 
are accelerated and even ejected from the cluster. However, more calculations 
are required in order to estimate the net amount of outflowing matter.
A detailed comparison of our models with multi-wavelength observations will
also be the subject of a forthcoming communication.

\begin{acknowledgements} We thank our anonymous referee for central comments
and suggestions. This study has been 
partly supported by grants AYA2004-08260-C03-01 and AYA 2004-02703
from the Spanish Ministerio de Educaci\'on y Ciencia (Spain),  
and Conacyt (M\'exico) grant 47534-F. We also acknowledge the
Institutional Research Plan AV10030501 of the Astronomical Institute, 
Academy of Sciences of the Czech Republic and project LC06014 Center for 
Theoretical Astrophysics.

\end{acknowledgements}


\begin{thebibliography}

\bibitem{1} Cant{\'o}, J., Raga, A.C. \&  Rodr\'\i{}guez, L.F. 2000, 
            ApJ, 536, 896
\bibitem{2} Chevalier, R. A. \& Clegg, A. W., 1985, Nature, 317, 44
\bibitem{3} Leitherer, C., Schaerer, D., Goldader, J.D. et al. 1999,
             ApJS, 123, 3
\bibitem{4} Melioli, C. \& de Gouveia Dal Pino E.M. 2004,  
            A\&A, 424, 817
\bibitem{5} Plewa, T. 1995, MNRAS, 275, 143
\bibitem{6} Raga, A.C., Vel{\'a}zquez, P.F., Cant{\'o}, J.,
             Masciadri, E. \& Rodriguez, L.F. 2001, ApJ, 559, L33
\bibitem{7} Rockefeller, G., Fryer, C.L., Melia, F. \& Warren, M.S.
            2004 ApJ, 604, 662
\bibitem{8} Rockefeller, G., Fryer, C.L., Melia, F. \& Wang, D.
            2005, ApJ, 623, 171
\bibitem{9} Stevens, I.R. \& Hartwell, J.M. 2003, MNRAS, 339, 280
\bibitem{10} Silich, S., Tenorio-Tagle G. \&
             Mu\~noz-Tu\~n\'on, C. 2003, ApJ, 590, 796 
\bibitem{11} Silich, S., Tenorio-Tagle, G., Rodr\'\i guez-Gonz\'alez, A. 
             2004, ApJ, 610, 226
\bibitem{12} Silich S., Tenorio-Tagle, G., Mu\~noz-Tu\~n\'on, C. \& 
             Palou\v{s}, J., 2006,
             IAU Symposium 237 ``Triggered Star Formation in a Turbulent ISM''
             (in press)
\bibitem{13} Tenorio-Tagle, G., Silich, S., Rodr\'iguez-Gonz\'alez A. 
             \& Mu\~noz-Tu\~n\'on, C., 2005a, ApJ, 620, 217
\bibitem{14} Tenorio-Tagle, G., Silich, S., Rodr\'iguez-Gonz\'alez A. 
             \& Mu\~noz-Tu\~n\'on, C., 2005b, ApJ, 628, L13
      
\end{thebibliography}
\end{document}